\documentclass[twocolumn,showpacs,preprintnumbers,amsmath,amssymb]{revtex4}
\pdfoutput=1
\usepackage{amsmath}
\usepackage{graphicx}
	\graphicspath{{pra/}}
\usepackage{mymacros}

\newcommand{\void}[1]{\ignorespaces}

\begin{document}

\title{Entanglement Resonance in Driven Spin Chains}

\author{Fernando Galve}
\author{David Zueco}
\author{Sigmund Kohler}
\author{Eric Lutz}
\author{Peter H\"anggi}
\affiliation{Institut f\"ur Physik, Universit\"at Augsburg,
        Universit\"atsstra{\ss}e~1, D-86135 Augsburg, Germany}

\date{\today}

\begin{abstract}
We consider a spin-$1/2$ anisotropic XY model with time-dependent
spin-spin coupling as means of creating long-distance
entanglement.  We predict the emergence of significant entanglement
between the first and the last spin whenever the ac part of the
coupling has a frequency matching the Zeeman splitting. In particular,
we find that the concurrence assumes its maximum with a vanishing dc
part.  Mapping the time-dependent Hamiltonian within a rotating-wave
approximation to an effective static model provides qualitative and
quantitative understanding of this entanglement resonance. Numerical
results for the duration of the entanglement creation and its length
dependence substantiate the effective static picture.
\end{abstract}

\maketitle

\section{Introduction}
Entanglement is a key resource for many quantum information and
computation protocols, such as teleportation \cite{teleport},
superdense coding \cite{densecode}, and cryptography \cite{crypto}. 
The successful storage and transfer of quantum information requires
effective mechanisms to create entangled states over large distances.
Since entanglement is generated mostly by local interactions, it is
initially short-ranged and, thus, has to be distributed via quantum
channels. Lately it has been noticed that spin
chains are promising candidates for this task \cite{bose03}. Various
spin-spin interactions, like e.g.\ Ising or Heisenberg coupling, have
been considered for entanglement creation, and their static as well as
their dynamical properties have been investigated \cite{amico2004,
fitz06, plas07, cubi08, bruder2006}.  Spin chains thus turned out to
be efficient quantum channels for controlled entanglement
distribution.

A particular spin chain is the quantum anisotropic XY model.
Irrespective of the magnitude of the anisotropy, it can be solved
exactly with the help of a Jordan-Wigner transformation and therefore
became a paradigmatic model in many-body physics \cite{XY1961}.
In the context of quantum information \cite{amico2008}, its
experimental implementation with optical lattices
\cite{lewenstein2003}, quantum dots \cite{qdots}, and Josephson
junctions \cite{joseph} has been proposed.

Thus far, most studies consider transfer of entanglement rather than its
generation \cite{Bosereview2007}, or its presence in systems with
static interactions \cite{venuti2006a, Ferreira2008,Difranco2008}.
In this paper, by contrast, we analyze a spin-$1/2$ XY
chain with periodically \textit{time-dependent} nearest-neighbor
coupling with separable initial state and find entanglement
\textit{creation} between the first and
the last spin of the chain.  As it may be difficult to access
individual spins in a controlled manner, we restrict ourselves to
chains with global time-dependent spin-spin coupling. 
%
%By studying the response of the system as a function of the
%driving frequency, we predict entanglement resonance.
Remarkably, entanglement created in that way turns out to be
significantly larger than the one in related static systems
\cite{rings, osterloh}.  We gain further insight by mapping the
time-dependent spin chain to a static model, which also provides
information on the length dependence and the duration of the
entanglement creation. 

%%%%%%%%%%%%%%%%%%%%%%%%%%%%%%%%%%%%%%%%%%%%%%%%%%%%%%%%%%%%%
\section{Spin chain Hamiltonian}
The anisotropic XY model in a transverse field $B$
and with time-dependent nearest-neighbor coupling $J(t)$ is described
by the Hamiltonian form (we put $\hbar=1$)
%%%%%%%%%%%%%%%%%%%%
\begin{equation}
H=\frac{B}{2}\!\sum_{n=1}^N\sigma_n^z+\frac{J(t)}{4}\!\sum_{n=1}^{N-1}
\Big[
(1+\gamma)\sigma_n^x\sigma_{n+1}^x+(1-\gamma)\sigma_n^y\sigma_{n+1}^y
\Big] ,
\label{H1}
\end{equation}
%%%%%%%%%%%%%%%%%%%%
where the $\sigma$'s are the usual Pauli matrices and $\gamma$
denotes the anisotropy parameter.
We focus on situations in which the coupling strength $J(t)$ is smaller than
the field strength $B$ and are interested in the spectral response of the
chain when the coupling is periodically modulated. All other
parameters have arbitrary but fixed values.
We also suppose that the spins are initially uncoupled, $J(t)=0$ for
$t<0$, and cooled down to the fully-aligned separable state 
%$ 
\begin{equation}
\label{psi0}
\ket{\psi(t{=}0)} = \ket {0000\ldots} \, ,
\end{equation}
%$ 
which is the ground state of the Hamiltonian \eqref{H1} with $J=0$.
At $t=0$, we switch on a coupling consisting of  a dc contribution
$J_0$ and a sinusoidal ac part with amplitude $J_1$,
\begin{equation} 
\label{jota}
J (t>0) = J_0+J_1\sin (\omegad t) \,.
\end{equation}
In the limit $J_1 \to 0$, the coupling suddenly switches to a constant
value, while for $J_1\neq 0 $, we are able to probe the
frequency-dependent response of the system. 
We quantify entanglement between the two ends of the chain with the
help of the concurrence $C=\max
\{\lambda_1-\lambda_2-\lambda_3-\lambda_4,0\}$. The $\lambda$'s are
the ordered square roots of the eigenvalues of $\rho(\sigma_y^1
\otimes \sigma_y^2) \rho^* (\sigma_y^1 \otimes \sigma_y^2)$
with $\rho$ being the reduced density matrix of the two spins
\cite{conc}.

%----------------
\begin{figure*}
\includegraphics{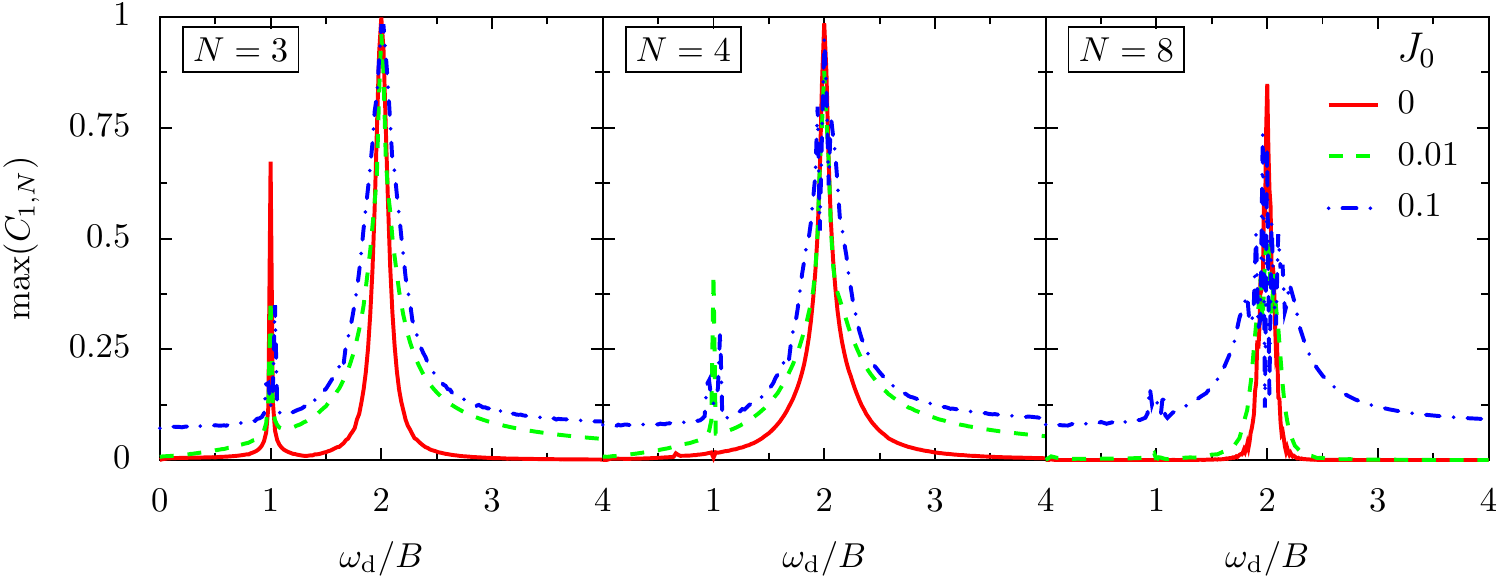}
\caption{(color online) Maximum concurrence obtained in a time
window up to $4N/{\rm{max(J_0,J_1)}}$, between spins $1$ and
$N$ for different frequencies and chain lengths $N$ with $J_1=0.1 B$,
$\gamma=1$, and $J_{0}=0$ (solid), $0.01 B$ (dashed), $0.1 B$
(dash-dotted).  The effective anisotropy thus has the values
$\tilde\gamma = \gamma J_1/2J_0 =\infty$, $5$, $0.5$.}
\label{fig:fig1}
\end{figure*}%

%----------------
\section{Entanglement resonance}
By direct numerical integration, we investigated the time
evolution of the concurrence between the first and the last spin for
different driving frequencies $\omegad$, chain lengths $N$, and
parameters $\gamma$, $J_1$, $J_0$. We determined the maximal
concurrence in the time interval $[0,\ldots,4 N/\max(J_0,J_1)]$.
The results shown in Fig.~\ref{fig:fig1} reveal that at $\omegad =
2 B$, irrespective of the other parameters, the concurrence assumes
during that time interval a value close to unity and is significantly
larger than for other
frequencies---we term this \textit{entanglement resonance}.  Height
and width of the resonance peak depend on the intensities $J_0$ and
$J_1$ and on the chain length $N$; see Fig.~\ref{fig:fig2}.  We
also notice the existence of a much smaller secondary peak at
$\omegad=B$. Contrary to the main peak, its amplitude strongly
decreases with decreasing coupling intensity and increasing chain
length.  We henceforth focus on the peak at $\omegad = 2B$.

\subsection{Rotating-wave approximation (RWA)}
Deeper understanding of the observed entanglement resonance
can be gained by analyzing the time-dependent
Hamiltonian~\eqref{H1} within rotating-wave approximation.
Since entanglement properties of a system are not changed by local
unitary operations on individual subsystems, it is convenient to
transform the XY Hamiltonian to the interaction picture, $\widetilde H
= \exp(i H_0 t) H \exp(-i H_0 t)$, with $H_0=(B/2)
\sum_i\sigma_i^z$.
By introducing the shift operators
$\sigma^\pm=\frac{1}{2}(\sigma^x\pm i\sigma^y)$, we obtain
\begin{equation}
\label{eq3}
\begin{split}
\widetilde H(t)
= \frac{J(t)}{2}  & \sum_{n=1}^N \Big[
    \sigma_{n}^+\sigma_{n+1}^- 
+
    \sigma_{n+1}^+\sigma_{n}^- 
\\ 
 + &\gamma e^{2 i B t} \sigma_n^+\sigma_{n+1}^+
 + \gamma e^{-2 i B t} \sigma_n^-\sigma_{n+1}^-
 \Big] \,.
\end{split}
\end{equation}
The first two terms swap the states of
spins $n$ and $n+1$, while the last two terms pairwise create (destroy)
excitations, which here is the origin of entanglement generation. 
If the driving frequency obeys the resonance condition $\omegad = 2B$
and, moreover, is much larger than both $J_0$ and $J_1$, we can
within RWA replace the Hamiltonian~\eqref{eq3} by its time
average 
\begin{equation}
\widetilde {H}_{R} = 
\frac{J_{0}}{2}\sum_{n=1}^N \Big[
   \sigma_n^+\sigma_{n+1}^- 
 + \tilde \gamma \sigma_n^+\sigma_{n+1}^+ + \text{H.c.}
\Big]
\label{effective}
\end{equation}
with the effective anisotropy $\tilde \gamma =\gamma J_1/2J_0$. 
This means that for resonant driving, the time dependent XY
model \eqref{H1} can be mapped to the static XY model~\eqref{effective}
without any Zeeman field.  In both cases, the entanglement generated
between the two end spins is maximal and controlled by the parameter
$\widetilde \gamma$ and the chain length $N$. Note
that $J_0\to 0$ corresponds to the infinitely anisotropic limit
$\tilde\gamma\to\infty$.
\begin{figure}
\includegraphics[scale=0.8]{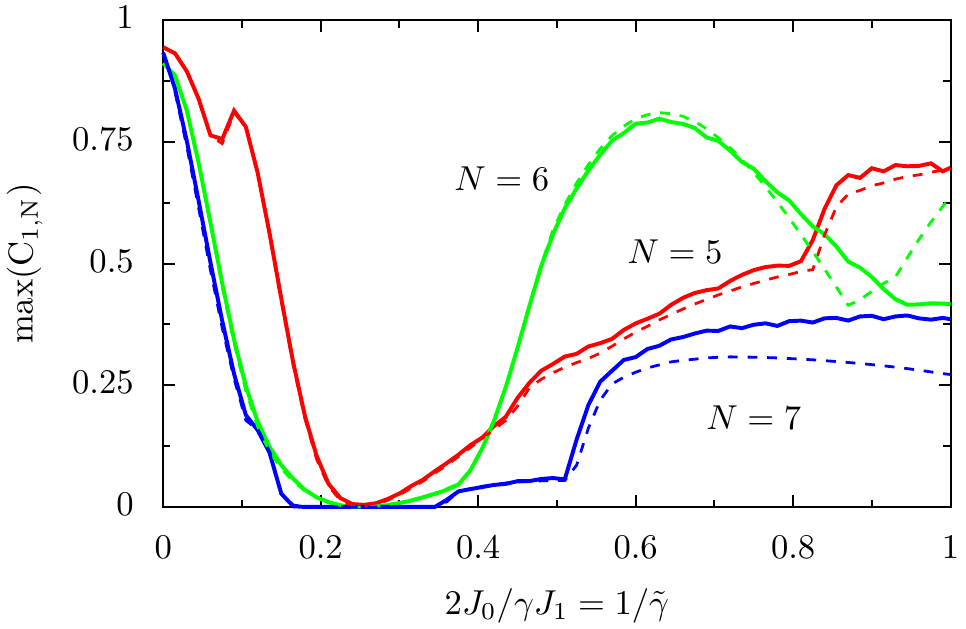}
\caption{(color online) Maximum obtained concurrence between the end spins for
$\omegad=2B$ as a function of the dc interaction $J_{0}$ for various
chain lengths, $J_1=0.1 B$ and $\gamma=1$.  The solid lines are
obtained with the full time-dependent Hamiltonian~\eqref{H1}, while
the dashed lines mark the RWA solution.}
\label{fig:fig2}
\end{figure}
Figure~\ref{fig:fig2} shows the concurrence between the end spins
as a function of the anisotropy parameter $\tilde \gamma$ for resonant
driving.  Two important points are worth being mentioned: First, the
concurrence approaches unity in the limit of vanishing $J_0$, i.e.\
for infinite $\tilde \gamma$. In this limit, the amount of entanglement no
longer depends on $J_1$ and $\gamma$.  Second, the agreement of the
exactly evaluated concurrence and the RWA solution is excellent, which
demonstrates that RWA is appropriate.
Moreover, Fig.~\ref{fig:fig3} shows that this approximation also
captures the entanglement dynamics, besides some small oscillations
stemming from neglected rapidly oscillating terms.
%--------------
\begin{figure}[t]
\includegraphics[scale=0.8]{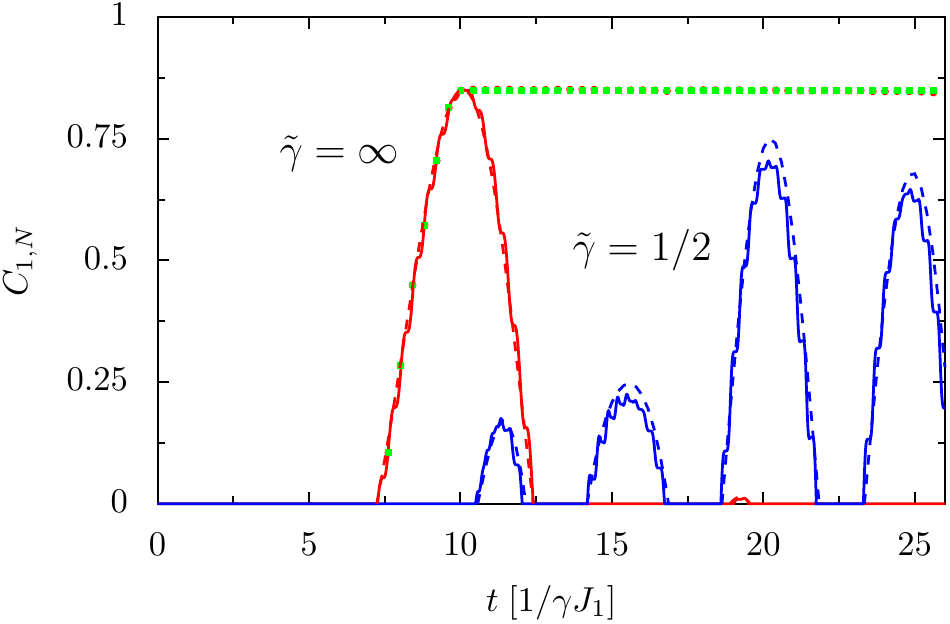}
\caption{(color online)
Entanglement dynamics for a chain of $8$ spins with driving frequency
$\omegad = 2 B$ for two different effective anisotropies
$\tilde\gamma = \gamma J_1/2J_0$. As in Fig.~\ref{fig:fig1},
$J_1=0.1 B$, $\gamma=1$.  Solid lines mark the exact numerical solution,
while the dashed lines are computed within RWA.
The symbols mark the time evolution for switching off the driving
(squares) and for changing the frequency to $\omega = 3B$ (circles) after the
concurrence maximum is reached at time $t_\text{arrival}$. Both
results cannot be distinguished for the chosen resolution.
}
\label{fig:fig3}
\end{figure}

%%%%%%%%%%%%%%%%%%%%%%%%%%%%%%%%%%%%%%%%%%%%%%%%%%%%%%%%%%
\section{Entanglement dynamics for resonant driving}
In order to investigate the entanglement dynamics, we consider the
exact time-evolution and discuss it within RWA.  In doing so, we find
the value of the effective anisotropy parameter $\tilde\gamma$
determines the qualitative behaviour.

\subsection{Strong anisotropy}

For a three-spin chain in the limit $\tilde\gamma\to\infty$ $(J_0=0)$, the repeated action of the Hamiltonian~\eqref{effective}
on the initial state creates the cyclic sequence $\ket{000}
\rightarrow \ket{110}+\ket{011} \rightarrow \ket{000}$.  This implies
that the quantum dynamics is a coherent oscillation between only these two
states.  The corresponding concurrence reads $C_{1,3}=|\sin(\gamma J_1/2 \sqrt{2} t)|$.  In particular, at
certain times, spins 1 and 3 are fully entangled, $C_{1,3}=1$.  The
exact time-evolution (not shown) agrees very well with the RWA prediction.
The three-spin case also reveals the difference between an open and a
closed chain: For the closed chain, which is translation-invariant,
the fully entangled state $|110\rangle+|011\rangle$ would be replaced
by $\ket{110} +\ket{011} +\ket{101}$ which has lower bipartite concurrence.
This emphasizes that lack of translation invariance supports the
entanglement creation between the ends of the chain.

For longer chains, the situation becomes more involved, but still can
be understood qualitatively.  Because the Hamiltonian conserves parity
and the initial state has zero excitations, the system will remain at
all times in a subspace of states having an even number of
excitations. This together with the fact that the chain is open can be
used to argue why at resonance there is such a huge amount of
entanglement.  Further, this argument also leads to the conclusion
that at the point of maximum entanglement the reduced state of spins
in the ends of the chain is $(\ket{00}+\ket{11})/\sqrt{2}$ for even
chains and $(\ket{01}+\ket{10})/\sqrt{2}$ for odd chains, plus a mixed
state contribution which is smallest the highest the concurrence. A
more detailed argumentation can be read in the Appendix.

The resulting entanglement dynamics is shown in Fig.~\ref{fig:fig3}: 
We find that the concurrence begins to grow after a given time, and
reaches a maximum value at a time $t_\text{arrival}$.
Thereafter, it decays.  However, two ways of maintaining the achieved
concurrence come to mind:  One can either simply switch off the
driving, i.e.\ $J(t>t_\text{arrival})=0$ or shift the driving frequency
to an off-resonant value $\omega_d\neq2B$.
The dotted lines in Fig.~\ref{fig:fig3} show that both strategies
freeze the entanglement as desired.  This certainly requires knowledge
of $t_\text{arrival}$ which behaves very regularly and can
be well estimated, as we demonstrate below.  Moreover, switching off
the driving parameters has to be much faster than the typical time
scale of the system, as we assume throughout this work.  The same
applies also to the onset of the driving.

\subsection{Moderate anisotropy}

For finite anisotropy $\tilde\gamma$ ($J_0\neq0$), the dynamics
becomes rather complex, see Fig.~\ref{fig:fig3}. The concurrence
assumes several local maxima until the highest one is reached.
Moreover, we find that the concurrence maximum became lower. This is due
to the presence of swapping terms, which spoil the argumentation of
the appendix.  These terms basically will mix the subsets
$\{\ket{00},\ket{11}\}$ and $\{\ket{01},\ket{10}\}$ 
and thus reduce the maximum achievable amount of concurrence.

Thus, we can conclude that the anisotropic limit $\tilde\gamma \to \infty$
($J_0=0$) is the optimal working point and, henceforth, restrict our
discussion to this limit.

%%%%%%%%%%%%%%%%%%%%%%%%%%%%%%%%%%%%%%%
\section{Scalability and arrival time}
Our next goal is to find the arrival time $\ta$ and the corresponding
concurrence maximum as a function of the chain length $N$.
Direct integration of the time-dependent Schr\"odinger
equation was only possible for up to $12$ spins.  The solution for 8
spins, however, already demonstrates that for $\omegad=2B$, the RWA
Hamiltonian~\eqref{effective} captures the global behavior very well,
see Fig.~\ref{fig:fig3}.  Therefore, we can make even further progress
by mapping the RWA Hamiltonian to a model for which an exact solution
is known.  For the unitary transformation
$\mathcal{S}=\prod_{i=1,3,\ldots}\sigma_i^x$ which flips the spins
with odd site number, we find the duality relation
\begin{equation}
\tilde\gamma \tilde{H}_{\tilde\gamma=0}
= \mathcal{S} \, \tilde{H}_{J_0=0} \, \mathcal{S}^\dagger .
\label{map}
\end{equation}
This means that the Hamiltonian for the infinitely anisotropic case
can be cast as a scaled isotropic Hamiltonian, while our initial state
is mapped to the Neel state: $|1010\ldots\rangle =
\mathcal{S}|0000\ldots\rangle$.  The Hamiltonian $\widetilde
H_{\gamma=0}$ can be diagonalized after a Jordan-Wigner transformation
\cite{mikeska1977}.

\begin{figure}
\includegraphics[scale=0.8]{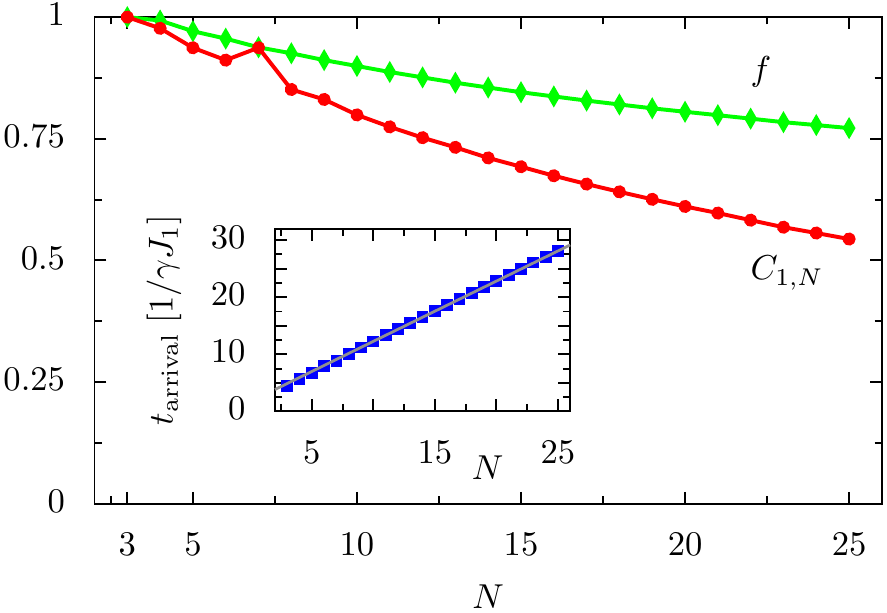}
\caption{(color online) 
Length dependence of the first maximum of the concurrence between
ends, $C_{1,N}(t)$,
and the corresponding fully entangled fraction $f$ for the RWA
Hamiltonian \eqref{effective} with $J_{0}=0$.  The inset shows the
length dependence of the arrival time at which the
entanglement assumes its maximum.
\label{fig:fig4}}
\end{figure}

Figure~\ref{fig:fig4} shows that the maximum entanglement achieved
decreases with the chain length rather slowly.  For very short chains we
find almost perfect entanglement, as predicted above within RWA, while
for length $N=25$, the concurrence still possesses the appreciable
value $C_{1,25}\approx 0.5$.  
A typical figure of merit in communication protocols is the ``fully
entangled fraction'', defined as $f=\mathop{\mathrm{max}}\langle
e|\rho|e\rangle$, where $\{|e\rangle\}$ is the set of all maximally
entangled states \cite{f}.  Quantum communication protocols are
superior to their classical counterparts whenever this fraction is
higher than $2/3$. In figure~\ref{fig:fig3} it is seen that this
magnitude greatly surpasses the classical efficiency for rather long
chains.

Note that a chain of length $N=7$ represents a particular case in
which the concurrence equals the fully entangled fraction.
We cannot provide an intuitive explanation for this
anomalous behavior.
%--------------
%--------------

Already above, we mentioned the importance of knowing the time $\ta$
at which the concurrence assumes its maximum.  In the strongly
anisotropic limit $\tilde\gamma\to\infty$ ($J_0=0$), we can provide a
good estimate for the arrival time with the following reasoning:
A typical local excitation will be transported with group velocity
$v_k = d\epsilon_k/dk$, where for $J_0=0$, the eigenenergies $\epsilon_k =
(\gamma J_1/2)\cos(k)$ are determined by the wave number $k=\pi
m/(N+1)$, $m=1,\ldots,N$ and form a band.  Since the initial state
$|0000\ldots\rangle$ is located in the center of the band, the relevant
wave number is $k\approx \pi/2$.  Thus, the time scale for traversing
the chain is 
\begin{equation}
t^*= \frac{N}{v_{\pi/2}} =  \frac{2N}{\gamma J_1}.
\end{equation}
The inset of Fig.~\ref{fig:fig3} shows that 
\begin{equation}
\ta \approx \frac{1.7}{\gamma J_1}+\frac{t^*}{2}=\frac{ 1.7 + N}{\gamma J_1},
\end{equation}
i.e., it grows linearly with the chain length.  The factor $1/2$ on
$t^*$ reflects the fact that counter-propagating excitations will meet
already in the middle of the chain thus establishing distant
entanglement.

Recently, Wichterich and Bose \cite{bose} computed the fully
entangled fraction in spin chains with isotropic nearest-neighbor
interaction.  Starting from the mixed Neel state
$(1/2)|0101\ldots\rangle\langle0101\ldots|
+(1/2)|1010\ldots\rangle\langle1010\ldots|$, they found that switching
on a constant interaction entangles the spins located at the
end sites.  The unitary transformation \eqref{map} maps this
model to the limit $\tilde\gamma\to\infty$ of the RWA Hamiltonian
\eqref{effective}.  Moreover, our discussion of the entanglement
dynamics within RWA vividly explains why in their case the isotropic
model permits the creation of a remarkably high entanglement.

%%%%%%%%%%%%%%%%%%%%%%%%%%%%%%%%%%%%%%%%%%%%%%%%%%%%
\section{Implementation with optical lattices}
The realization of a XY chain with anisotropy $\gamma=1$ has been
proposed for experiments with cold atoms in a one-dimensional optical
lattice that in transverse direction forms a bistable potential
\cite{lewenstein2003}.  The ground-state doublet of each double-well
forms the ``spin'' degree of freedom.  Then our initial state
\eqref{psi0} corresponds to a Mott-insulator state, which has already
been realized experimentally \cite{bloch2002}.  There the tunnel
barriers in longitudinal direction can be up to $\sim22E_\mathrm{r}$,
where the recoil energy $E_\mathrm{r}$ typically lies in the kHz
regime.  This is more than sufficient for suppressing longitudinal
tunneling, such that each double well remains occupied with a single
atom, while the spin-spin interaction is given by a Bose-Hubbard
repulsion term. The repulsion term
is caused by an overlap of Wannier functions describing neighbouring atoms.
The amount of overlap is given by the barrier height, which can be
controlled and modulated via the laser intensity, yielding a time
dependent $J(t)$.   Thus a high barrier effectively yields no overlap
and hence no spin-spin interaction ($J=0$), whereas a low barrier can
yield values $J\sim 0.1\,\mathrm{kHz}$ \cite{duan2003}.  The Zeeman
field $B$ corresponds to the tunnel splitting of the double-well
potential and is of the order $0.1E_\mathrm{r}$ \cite{morsch2006}.
Though, it can be manipulated by changing the depth of the double well
potential, so that $B$ is greater than, but of the order of, $J(t)$.
This implies that the switching times of the Zeeman fields have to be
considerably smaller than $1\,\mathrm{ms}$.  Coherence times for atoms
in such optical lattices can be much larger than the system time scale
and, thus, decoherence should not play a major role. Moreover, the
initial state $\ket{000...0}$ can be imposed by tayloring the field
$B$ to be much higher than thermal excitation energy $k_BT$ due to the
environment.  Finally, the ``spin state'' in the transverse double
well can be probed by fluorescence measurement of the atoms.

\section{Conclusions}
We have shown that proper ac driving can induce almost perfect
entanglement between the first and the last spin of an anisotropic XY
chain. As a most striking feature, we found that the driven chain
bears the potential for a considerably larger entanglement than the
formerly studied static chains.
We identified a resonance condition which leads to maximal
entanglement and also provide a reliable estimate for the time after
which this entanglement is reached.  The latter is crucial for
freezing the entanglement once it is created.  Our analysis within a
rotating-wave approximation contributed to a qualitative and
quantitative understanding of how the entanglement is built up:
pairwise flipping of neighboring spins of an open chain favors
correlations between the end spins.
Moreover, we found that the maximum entanglement decreases only weakly
with the chain length, while the entanglement is built up during a
time that is linearly length dependent.  Thus our protocol demonstrates good
scalability which is a major requirement for the implementation of
quantum communication protocols.
A natural application of our scheme is quantum communication via state
teleportation.  This is possible because the fully entangled fraction
between the first and the last spin is sufficiently large, such that a
spin singlet can be purified \cite{f}.
Let us finally emphasize that our protocol can be implemented with
three different experimental setups, namely an anisotropic chain with
sinusoidal driving, an infinitely anistropic chain with a sudden
switch and an isotropic chain with initial Neel state. This provides a
broad choice for its application.

\section{Acknowledgements}

We gratefully acknowledge support by the German Excellence Initiative
via the ``Nanosystems Initiative Munich (NIM)'', as well as by DFG
through SFB 484, SFB 631, and the Emmy Noether program (LU1382/1-1).

%-------------------------------------------------------
\appendix
\section{Entanglement in strongly anisotropic chains}

Due to the parity preserving character of the Hamiltonian and the
fact that our initial state $\ket{00...0}$ has a definite parity, the
reduced density matrix of the spins at the ends of the chain $\rho_R$
is of the form $p_1\ket{00}\bra{00} +p_2(\ket{01}\bra{01}
+\ket{10}\bra{10}) +p_3\ket{11}\bra{11} +(\alpha\ket{00}\bra{11}
+\beta\ket{01}\bra{10} +\text{H.c.})$, i.e., in the basis
$\{|00\rangle, |01\rangle, |10\rangle, |11\rangle\}$, it reads
\begin{equation}
\rho_R=\left(\begin{array}{cccc}
p_1&0&0&\alpha\\
0&p_2&\beta&0\\
0&\beta^*&p_2&0\\
\alpha^*&0&0&p_3
\end{array}\right) ,
\end{equation}
and the corresponding concurrence is
\begin{equation}
C=2\max(0,|\alpha|-p_2,|\beta|-\sqrt{p_1p_3}).
\end{equation}
Because the chain is open and the Hamiltonian flips spins pairwise
at adjacent sites, we find $\beta=0$ for even
chains and $\alpha=0$ for odd chains. 

For even chains, the proof of this statement is as follows: The term
$\ket{01}\bra{10}$ stems from evaluating the trace over density
operators of the form $\ket{0[x]1}\bra{1[y]0}$, where the blocks $[x]$
and $[y]$ represent the rest of the chain.  Obviously, only terms with
$[x]=[y]$ yield a non-vanishing contribution.  We demonstrate by
\textit{reductio ad absurdum} that it is impossible to fulfill this
condition: Let us assume that states $\ket{0[x]1}\bra{1[x]0}$ can
occur. By applying the pairwise flipping Hamiltonian we have
$[x]=[y]1$ for the ket and $[x]=1[z]$ for the bra, where the blocks
$[y]$ and $[z]$ are yet one spin shorter and, thus, consist of an odd number
of spins. Hence $[x]=1[x']1$, such that the operator becomes
$\ket{01[x']11}\bra{11[x']10}$. Again, by the same reasoning we find
the requirement $[x']=1[y']$ for the ket and $[x']=[z']1$ for the bra.
Therefore $[x']=1[x'']1$ and, thus,
$\ket{011[x'']111}\bra{11[x']10}$. Repeating this procedure, we
end up with a collection of ever smaller blocks
$[x],[x'],[x''],....,[x^{(n)}]$ all of which possessing an even number of spins.
Eventually, we remain with the operator
$\ket{011...1[x^{(n)}]1...1}\bra{11...1[x^{(n)}]1...10}$. From the ket
we find the condition that $[x^{(n)}]=[10]$, while from the bra
follows $[x^{(n)}]=[01]$ in
order to have a total even number of $1$s and thus conserve parity.
Thus we can conclude that initial hypothesis $[x]=[y]$ must be wrong.
This proves that for even chains $\beta=0$, and so the concurrence
reduces to $C=2\max(0,|\alpha|-p_2)$. This line of reasoning can be
adapted to the case of odd chains, for which one obtains $\alpha=0$.

Yet, in order to obtain a high concurrence, we need $|\alpha|\gg p_2$,
as we find in our numerical studies.  The trace condition for density
matrices yields $p_1+p_3=1-2p_2$, while positivity requires
$\alpha\leq\sqrt{p_1p_3}$. Note that for pure states, $\alpha =
\sqrt{p_1p_3}$. Thus, maximizing $\alpha$ necessarily requires $p_2$
be small, that is, if at any instance of time, $\alpha$
starts to increase, as it happens when $p_2$ becomes smaller,
the concurrence increases as well. Clearly this can occur only
at certain times, which is why we see entanglement peaks.

At times of maximum concurrence, the resulting state shared between
spins $1$ and $N$ is then $a_1\ket{00}+a_2\ket{11}$ for even chains
and $a_1\ket{01}+a_2\ket{10}$ for odd chains, where we have ignored a
small mixed state contribution.

\bibliography{spins}% Produces the bibliography via BibTeX.

\end{document}